\documentclass[prl,aps,twocolumn,floatfix,showpacs]{revtex4}
\usepackage{graphicx,graphics,psfrag,amsmath,calc}
\usepackage{epsfig}
\usepackage{color}
\begin{document}
\title{Evolution from BCS to BEC superfluidity in the
presence of spin-orbit coupling}
\author{Li Han and C. A. R. S{\'a} de Melo}
\affiliation{School of Physics, Georgia Institute of Technology,
Atlanta, Georgia 30332, USA}
\date{\today}

\begin{abstract}
We discuss the evolution from BCS to BEC superfluids in the presence
of spin-orbit coupling, and show that this evolution is just a
crossover in the balanced case. The dependence of several
thermodynamic properties, such as the chemical potential, order
parameter, pressure, entropy, isothermal compressibility and spin
susceptibility tensor on the spin-orbit coupling and interaction
parameter at low temperatures are analyzed. We studied both the case
of equal Rashba and Dresselhaus (ERD) and the Rashba-only (RO)
spin-orbit coupling. Comparisons between the two cases reveal
several striking differences in the corresponding thermodynamic
quantities. Finally we propose measuring the spin susceptibility as
a means to detect the spin-orbit coupling effect.

\pacs{03.75.Ss, 67.85.Lm, 67.85.-d}
\end{abstract}
\maketitle

%
%

Superfluidity is a ubiquitous phenomenon that is encountered in
nearly every area of physics including condensed matter, nuclear,
astro, and atomic and molecular physics. Superflow results from
strong correlations between particles, which for any given
interacting Fermi system could not be controlled externally until
recently with the advent of ultra-cold atoms. In standard condensed
matter there is a continuous search for new charged superfluids
(superconductors) since the type and strength of interactions can
not be tuned even within the same class of materials. In the case of
nuclear matter the issue of tunability of interactions is even
worse, being hopeless for neutron stars. However, the situation is
much more favorable for ultra-cold Fermi atoms, where the ability to
control interactions between particles, via Feshbach resonances, has
been demonstrated in experimental studies of the so-called crossover
from BCS to BEC superfluidity.

Further control of interactions is now possible through newly
developed experimental techniques that allow the production of
fictitious magnetic fields which couple to neutral bosonic
atoms~\cite{spielman-2009a, spielman-2009b}. These fictitious
magnetic fields are generated through an all optical process, but
produce real effects like the creation of vortices in the superfluid
state of bosons. Furthermore, artificial spin-orbit coupling has
also been produced in neutral bosonic systems~\cite{spielman-2011}
where the strength of the coupling can be controlled optically. In
principle the same techniques can be applied to ultracold
fermions~\cite{spielman-2011, chapman-sademelo-2011}, which, when
coupled with the control over the interaction using Feshbach
resonances, allows for the exploration of superfluidity not only as
a function of interactions, but also as a function of fictitious
magnetic fields~\cite{iskin-sademelo-2011}, or as a function of
spin-orbit coupling discussed here. An introduction to the effects
of controllable fictitious magnetic and spin-orbit fields can now be
found in the literature~\cite{dalibard-2011}.

It is in anticipation of experiments involving spin-orbit coupling
in fermionic atoms such as $^6$Li, $^{40}$K and isotopes
of Ytterbium, that we discuss here
the evolution from BCS to BEC superfluidity in the presence of
controllable spin-orbit couplings for balanced fermions
in three dimensions. We investigate
spin-orbit effects with
Dresselhaus~\cite{dresselhaus-1955} and/or Rashba~\cite{rashba-1984}
terms, and analyze several thermodynamic quantities
including the order parameter, chemical potential, thermodynamic
potential, entropy, pressure, isothermal compressibility,
and spin susceptibility tensor as a function of spin-orbit coupling
and interaction parameter at low temperatures.
We conclude that the BCS-to-BEC evolution for balanced fermions
including spin-orbit effects is just a crossover.

%
%

{\it Hamiltonian:}
To address the problem of the evolution from BCS to
BEC superfluidity in the presence of spin-orbit fields
for balanced or imbalanced Fermi-Fermi mixtures, we start
with the generic Hamiltonian density
\begin{equation}
{\cal H} ({\bf r})
=
{\cal H}_0 ({\bf r})
+
{\cal H}_I ({\bf r}).
\end{equation}
The single-particle Hamiltonian density is
\begin{equation}
\label{eqn:hamiltonian-single-particle} {\cal H}_0 ({\bf r}) =
\sum_{\alpha \beta} \psi^{\dagger}_{\alpha} ({\bf r}) \left[ {\hat
K}_{\alpha} \delta_{\alpha \beta} -h_i ({\bf
r})\sigma_{i,\alpha\beta} \right] \psi_{\beta} ({\bf r}),
\end{equation}
where $ {\hat K}_{\alpha} = - \nabla^2/(2 m_{\alpha}) - \mu_{\alpha}
$ is the kinetic energy in reference to the chemical potential
$\mu_{\alpha}$ , and $h_i ({\bf r})$ is the spin-orbit field along
the $i$-direction ($\alpha = \uparrow, \downarrow$, $i=x, y, z$).
The interaction term is
$
{\cal H}_I ({\bf r})
=
-g
\psi^{\dagger}_{\uparrow} ({\bf r})
\psi^{\dagger}_{\downarrow} ({\bf r})
\psi_{\downarrow} ({\bf r})
\psi_{\uparrow} ({\bf r}),
$
where $g$ is a contact interaction. In this paper we set $\hbar =
k_B = 1$.

%
%

{\it Effective Action:} The partition function at temperature $T$ is
$ Z = \int \mathcal{D}[\psi, \psi^\dagger] \exp \left(
 -S[\psi, \psi^\dagger]
\right) $ with action
\begin{equation}
\label{eqn:action-initial}
S[\psi, \psi^\dagger]
=
\int d\tau d {\bf r}
\left[
\sum_{\alpha}
\psi^\dagger_{\alpha} ({\bf r})
\frac{\partial}{\partial \tau}
\psi_{\alpha} ({\bf r}) +
{\cal H} ({\bf r}, \tau)
\right].
\end{equation}

Using the standard Hubbard-Stratanovich transformation that
introduces the pairing field
$
\Delta ({\bf r}, \tau)
=
g
\langle
\psi_{\downarrow} ({\bf r}, \tau)
\psi_{\uparrow} ({\bf r}, \tau)
\rangle
$
we can write the intermediate action
$ S_{\rm int}[ \psi, \psi^\dagger, \Delta, \Delta^\dagger ] = S_{\rm
no}[\psi, \psi^\dagger] + S_{\rm I}[ \psi, \psi^\dagger, \Delta,
\Delta^\dagger ], $ 
where the no-interaction action is
$$
S_{\rm no}[\psi, \psi^\dagger] = \int d\tau d {\bf r} \left[
\sum_{\alpha} \psi^\dagger_{\alpha} ({\bf r})
\frac{\partial}{\partial \tau} \psi_{\alpha} ({\bf r}) + {\cal H}_0
({\bf r}, \tau) \right],
$$
and the action due to the auxiliary field is
$$
S_{\rm I}
=
\int d\tau d {\bf r}
\left[
\frac{ \vert \Delta ({\bf r, \tau}) \vert^2 }
{g}
-
\Delta \psi^\dagger_{\uparrow} \psi^\dagger_{\downarrow}
-
\Delta^\dagger \psi_{\downarrow} \psi_{\uparrow}
\right].
$$

Using the four-dimensional vector
$ \Psi^\dagger = \{ \psi^\dagger_{\uparrow},
\psi^\dagger_{\downarrow}, \psi_{\uparrow}, \psi_{\downarrow} \}, $
the intermediate action becomes
$$
S_{\rm int} = \int d\tau d {\bf r} \left[ \frac{ \vert \Delta ({\bf
r, \tau}) \vert^2 } {g} + \frac{1}{2} \Psi^\dagger {\bf M} \Psi +
\frac{1}{2} ( \widetilde K_\uparrow + \widetilde K_\downarrow )
\right].
$$
The $4 \times 4$ matrix ${\bf M}$ is
\begin{equation}
\label{eqn:matrix-m}
{\bf M}
=
\left(
\begin{array}{cccc}
\partial_\tau + \widetilde K_\uparrow & - h_\perp & 0 & -\Delta \\
- h_\perp^* & \partial_\tau + \widetilde K_\downarrow &  \Delta & 0 \\
0 & \Delta^\dagger & \partial_\tau - \widetilde K_\uparrow &  h_\perp^* \\
-\Delta^\dagger  & 0 & h_\perp & \partial_\tau - \widetilde K_\downarrow
\end{array}
\right),
\end{equation}
where $h_{\perp} = h_x - i h_y$ corresponds to the transverse component
of the spin-orbit field, $h_z$ to the parallel component
with respect to the quantization axis $z$,
$\widetilde K_\uparrow  = {\hat K}_\uparrow - h_z$,
and $\widetilde K_\downarrow  = {\hat K}_\downarrow + h_z$.
Integration over the fields $\Psi$ and $\Psi^\dagger$ leads to
the effective action
\begin{equation}
S_{\rm eff}
=
\int d\tau d {\bf r}
\left[
\frac{ \vert \Delta ({\bf r, \tau}) \vert^2 }
{g}
-
\frac{T}{2V}
\ln \det \frac{{\bf M}}{T}
+
\widetilde K_+ \delta ({\bf r} - {\bf r}^\prime)
\right],
\end{equation}
where
$
\widetilde K_{+}
=
( \widetilde K_\uparrow + \widetilde K_\downarrow )/2.
$
%

%
%

{\it Saddle Point Approximation:} To proceed we use the saddle point
approximation $ \Delta ({\bf r}, \tau) = \Delta_0 + \eta ({\bf r},
\tau), $ and separate the matrix ${\bf M}$ into two parts. The first
one is the saddle point matrix ${\bf M}_0$, where the transformation
$\Delta ({\bf r}, \tau) \to \Delta_0$ takes ${\bf M} \to {\bf M}_0$.
The second one is the fluctuation matrix ${\bf M}_{{\rm F}} = {\bf
M} - {\bf M_0}$, which depends only on $\eta ({\bf r}, \tau)$ and
its Hermitian conjugate.

Using the saddle point approach we write the effective action as
$S_{\rm eff} = S_0 + S_{\rm F}$, where
$$
S_0
=
\int d\tau d{\bf r}
\left[
\frac{\vert \Delta_0 \vert^2}{g}
-
\frac{T}{2V} \ln \det \frac{{\bf M}_0}{T}
+
\widetilde K_{+} \delta ({\bf r} - {\bf r}^\prime)
\right]
$$
is the saddle point action and
$$
S_{\rm F}
=
\int d\tau d{\bf r}
\left[
\frac{\vert \eta ({\bf r}, \tau) \vert^2}{g}
-
\frac{T}{2V}
\ln \det
\left(
{\bf 1} + {\bf M}_0^{-1} {\bf M}_{\rm F}
\right)
\right]
$$
is the fluctuation action in all orders in the fluctuation field.
The effects of fluctuations at both zero temperature and near the
critical temperature will be discussed later.

A transformation to the momentum-frequency coordinates $({\bf k}, i
\omega_n)$, where $\omega_n = (2n + 1) \pi T$, leads to
$$
S_0
=
\frac{V}{T}
\frac{\vert \Delta_0 \vert^2}{g}
-\frac{1}{2}
\sum_{k, i\omega_n, j}
\ln
\left[
\frac{i\omega_n - E_j ({\bf k})}{T}
\right]
+
\sum_{\bf k}
\frac{{\widetilde K}_{+}}{T},
$$
where $E_j ({\bf k})$ are the eigenvalues of the matrix
\begin{equation}
{\bf H}_0
=
\left(
\begin{array}{cccc}
\widetilde K_\uparrow ({\bf k}) & - h_\perp ({\bf k}) & 0 & -\Delta_0 \\
- h_\perp^* ({\bf k}) & \widetilde K_\downarrow ({\bf k}) &  \Delta_0 & 0 \\
0 & \Delta_0^\dagger & -\widetilde K_\uparrow ({-\bf k}) &  h_\perp^* ({-\bf k}) \\
-\Delta_0^\dagger  & 0 & h_\perp (-{\bf k}) & - \widetilde K_\downarrow (-{\bf k})
\end{array}
\right),
\end{equation}
which describes the Hamiltonian of the elementary excitations in the
four-dimensional vector basis $ \Psi^\dagger = \left\{
\psi_{\uparrow}^\dagger ({\bf k}), \psi_{\downarrow}^\dagger ({\bf
k}), \psi_{\uparrow}(-{\bf k}), \psi_{\downarrow}(-{\bf k})
\right\}. $ The spin-orbit field is $ {\bf h}_\perp ({\bf k}) = {\bf
h}_R ({\bf k}) + {\bf h}_D ({\bf k}), $ where the first term is of
the Rashba-type $ {\bf h}_R ({\bf k}) = v_R \left( -k_y {\hat{\bf
x}} + k_x {\hat {\bf y}} \right), $ and the second is of the
Dresselhaus-type $ {\bf h}_D ({\bf k}) = v_D \left( k_y {\hat {\bf
x}} + k_x {\hat {\bf y}} \right). $ We assume, without loss of
generality, that $v_R > 0$ and $v_D > 0$. The magnitude of the
transverse field is then $ h_{\perp} ({\bf k}) = \sqrt{ \left( v_D -
v_R \right)^2 k_y^2 + \left( v_D + v_R \right)^2 k_x^2 }. $ In the
limiting cases of pure Rashba (R) with $v_D = 0$ and for equal
Rashba-Dresselhaus (ERD) couplings with $v_R = v_D = v/2$, the
transverse fields are $ h_{\perp} ({\bf k}) = v_R \sqrt{k_x^2 +
k_y^2}$ ($v_R > 0$) and $ h_{\perp} ({\bf k}) = v \vert k_x \vert $
($v > 0$), respectively.

%
%

{\it Order parameter and number equations:}
The saddle point thermodynamic potential
$\Omega_ 0 = T S_0$ is obtained by
integrating out the fermions leading to
$$
\Omega_0
=
V
\frac{\vert \Delta_0 \vert^2}{g}
-\frac{T}{2}
\sum_{{\bf k}, j}
\ln
\left\{
1 + \exp \left[ - E_j ({\bf k})/T \right]
\right\}
+
\sum_{\bf k}
{\bar K}_{+},
$$
with
$
{\bar K}_{+}
=
\left[
\widetilde K_\uparrow (-{\bf k})
+
\widetilde K_\downarrow (-{\bf k})
\right]/2.
$
The order parameter is determined via the minimization of $\Omega_0$
with respect to $\vert \Delta_0 \vert^2$ leading to
\begin{equation}
\label{eqn:order-parameter-general} \frac{V}{g} = -\frac{1}{2}
\sum_{{\bf k}, j} n_F \left[E_j ({\bf k}) \right] \frac{\partial E_j
({\bf k})}{\partial \vert \Delta_0 \vert^2},
\end{equation}
where
$ n_F \left[  E_j (\mathbf{k})  \right] = 1/(\exp\left[ E_j ({\bf
k})/T\right] + 1) $ 
is the Fermi function for energy $E_j ({\bf k})$. We replace the
contact interaction $g$ by the scattering length $a_s$ through the
relation $ 1/g = - m_+/(4\pi a_s) + (1/V) \sum_{\bf k} \left[
1/(2\epsilon_{{\bf k},+}) \right], $ where $ m_+ = 2 m_\downarrow
m_\uparrow / (m_\downarrow + m_\uparrow) $ is twice of the reduced
mass, $ \epsilon_{ {\bf k}, \alpha } = k^2 / (2m_\alpha) $ are the
kinetic energies, and $ \epsilon_{{\bf k}, +} = \left[ \epsilon_{
{\bf k}, \uparrow } + \epsilon_{ {\bf k}, \downarrow } \right] / 2.
$ The number of particles at the saddle point is obtained by $
N_{\alpha} = -
\partial \Omega_0
/
\partial \mu_{\alpha}
$, leading to
\begin{equation}
\label{eqn:number-general}
N_{\alpha}
=
\frac{1}{2}
\sum_{\bf k}
\left[
1 -
\sum_j n_F \left[ E_j ({\bf k}) \right]
\frac{\partial E_j ({\bf k})}{\partial \mu_{\alpha}}
\right].
\end{equation}
The self-consistent relations shown in
Eqs.~(\ref{eqn:order-parameter-general})
and~(\ref{eqn:number-general}) are general for arbitrary mass and
population imbalances. However, next, we particularize our
discussion to the case of a balanced system with equal masses.

%
%

{\it Balanced Populations:}
In the case of mass and population balanced systems,
the four eigenvalues of the matrix ${\bf H}_0$ are
$
E_1 ({\bf k})
=
\sqrt{
\left[
\varepsilon_1 ({\bf k})
\right]^2
+
\vert \Delta_0 \vert^2
},
$
$
E_2 ({\bf k})
=
\sqrt{
\left[
\varepsilon_2 ({\bf k})
\right]^2
+
\vert \Delta_0 \vert^2
},
$
$E_3 ({\bf k}) = - E_1 ({\bf k})$, and $E_4 ({\bf k}) = - E_2 ({\bf
k}).$ Here, the auxiliary energies are $ \varepsilon_1 ({\bf k}) =
\xi ({\bf k}) + h_{\perp} ({\bf k}), $ and $ \varepsilon_2 ({\bf k})
= \xi ({\bf k}) - h_{\perp} ({\bf k})$. The corresponding order
parameter equations at the saddle point level is
\begin{equation}
\label{eqn:order-parameter-balanced} \frac{V}{g} = \frac{1}{2}
\sum_{\bf k} \left[ \frac{X_1 ({\bf k})} { 2 E_1 ({\bf k})} +
\frac{X_2 ({\bf k})} { 2 E_2 ({\bf k})} \right],
\end{equation}
where $ X_m ({\bf k}) = \tanh \left[ E_m ({\bf k})/2T \right] $ ($m
= 1, 2$). Since the mixture of equal mass fermions is balanced, the
chemical potentials are the same $ \mu_{\uparrow} = \mu_{\downarrow}
= \mu, $ and the associated number equation is $N = - \partial
\Omega / \partial \mu$ that reduces to
\begin{equation}
\label{eqn:number-balanced}
N
=
\sum_{\bf k}
\left[
1
-
\frac{X_{1} ({\bf k})}{2E_1 ({\bf k})} \varepsilon_1 ({\bf k})
-
\frac{X_{2} ({\bf k})}{2E_2 ({\bf k})} \varepsilon_2 ({\bf k})
\right].
\end{equation}

\begin{figure} [htb]
\centering
\begin{tabular}{cc}
\epsfig{file=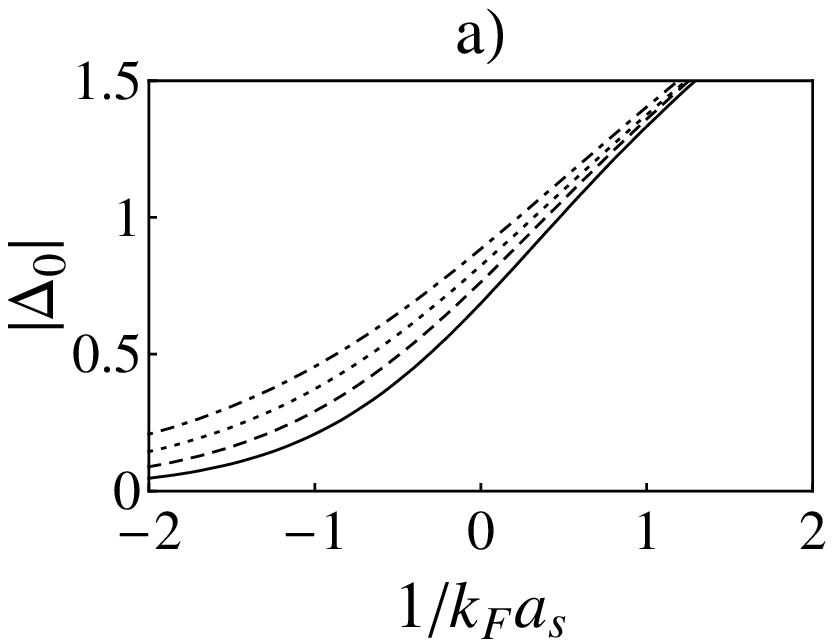,width=0.48 \linewidth} &
\epsfig{file=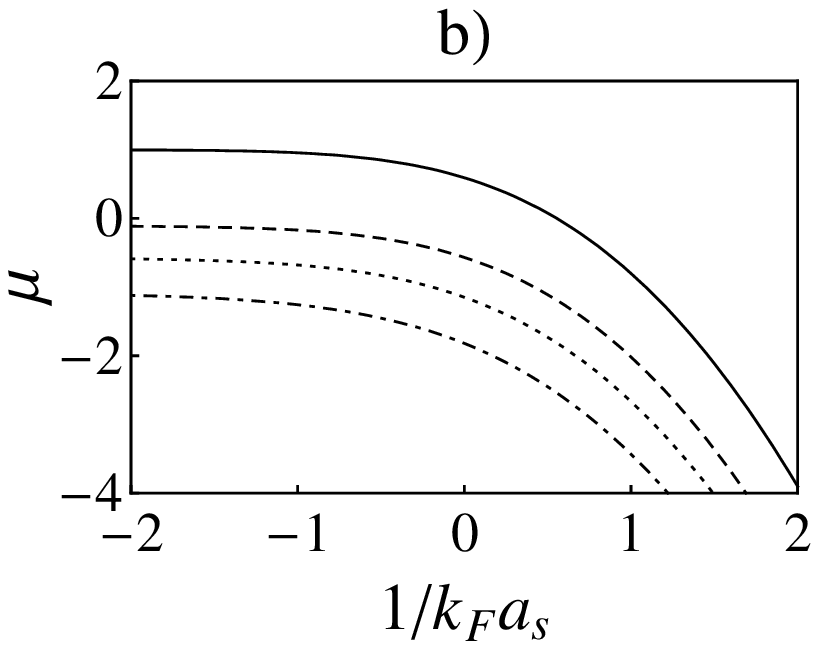,width=0.48 \linewidth}
\end{tabular}
\caption{ \label{fig:one} Order parameter $\vert \Delta_0 \vert$ and
chemical potential $\mu$ (in units of the Fermi energy $\epsilon_F$)
as a function of interaction parameter $1/(k_F a_s)$ for different
spin-orbit couplings $v_R/v_F = 0$ (solid), $v_R/v_F = 0.8$
(dashed), $v_R/v_F = 1.0$ (dotted), and $v_R/v_F = 1.2$ (dot-dashed)
at $T = 0$ in the RO case. Here $v_F = k_F/m$ is the Fermi velocity.
}
\end{figure}

In Fig.~{\ref{fig:one}}, we show the zero temperature behavior of
$\vert \Delta_0 \vert$ and $\mu$ as a function of $1/(k_F a_s)$ for
various values of spin-orbit coupling in the
equal-Rashba-Dresselhaus (ERD) and for Rashba-only (RO) cases. In
the ERD case the order parameter $\vert \Delta_0 \vert$ is
independent of $v$, and the chemical potential $\mu (v)$ is simply
$\mu (v) = \mu (0) - mv^2/2$, since the transverse field $h_{\perp}
({\bf k}) = v \vert k_x \vert$ can be eliminated by momentum shifts
along the $x$-direction, effectively gauging away spin-orbit effects
in the {\it charge or momentum} sector. This symmetry also implies
that the critical temperature $T_c$ as a function of $1/(k_F a_s)$
for finite $v$ is the same as that for $v = 0$. However, in the RO
case, shifts in momentum can not gauge away the spin-orbit coupling,
and $\vert \Delta_0 \vert $ increases with increasing $v_R$, while
$\mu$ decreases as $v_R$ increases, exhibiting the same tendency as
in the ERD case. In the BCS regime, the increase of $\vert \Delta_0
\vert$ with $v_R$ also leads to an increase of $T_c$ with increasing
$v_R$.

%

%
%

{\it Momentum distribution and excitation spectrum:} The momentum
distribution $n ({\bf k})$ is obtained from
Eq.~(\ref{eqn:number-balanced}) using the definition $N = \sum_{\bf
k} n ({\bf k})$. At fixed momentum component $k_z = 0 $ and fixed
interaction strength, the momentum distribution $n ({\bf k})$ shifts
continuously with increasing spin-orbit coupling in the BCS $\left[
1/(k_F a_s) \ll -1 \right]$ or unitarity regimes $\left[ 1/(k_F a_s)
\to 0 \right]$. For zero spin-orbit coupling, $n ({\bf k})$ is that
of a superfluid degenerate Fermi system with identical
single-particle bands $\xi ({\bf k})$ and has a nearly flat momentum
distribution until the Fermi momentum is reached. However, as the
spin-orbit coupling is turned on, non-identical single-particle
bands $\xi_{\Uparrow} ({\bf k}) = \xi ({\bf k}) - h_\perp ({\bf k})$
and $\xi_{\Downarrow}({\bf k}) = \xi ({\bf k}) + h_\perp ({\bf k})$
in the helicity basis $\vert{\Uparrow\rangle},
\vert{\Downarrow\rangle}$ emerge and produce a double structure with
a reasonably flat momentum distribution centered around finite
momenta in the $(k_x, k_y)$ plane. In the BEC regime $\left[ 1/(k_F
a_s) \gg 1 \right]$ the momentum distributions for weak and strong
spin-orbit coupling broadens substantially due to the loss of
degeneracy in the Fermi system when the chemical potential goes
below the minima of the helicity bands and becomes large and
negative. Even though there is a substantial change in the momentum
distribution as a function of the spin-orbit coupling, we notice
that the excitation energies $E_1 ({\bf k})$ and $E_2 ({\bf k})$ is
always gapped for all values of the interaction parameter $1/(k_F
a_s)$ or the spin-orbit field $h_\perp ({\bf k})$, immediately
suggesting that thermodynamic properties, which depend on the
excitation energies, evolve smoothly from the BCS to the BEC regime
in the balanced case for fixed values of spin-orbit coupling. The
omnipresence of a gap in the excitation spectrum shows that the
evolution from BCS to BEC superfluidity at finite spin-orbit
coupling for balanced systems is just a crossover. The situation is
different for imbalanced systems, where gapless regions emerge in
the excitation spectrum and topological phase transitions occur, so
long as the system is stable~\cite{iskin-sademelo-2006,
chuanwei-2011}. A thermodynamic signature of this crossover for
balanced systems is seen in the isothermal compressibility discussed
next.

%
%

{\it Isothermal compressibility:}
An important thermodynamic property, which can now be measured
experimentally using the fluctuation-dissipation theorem, is
the isothermal compressibility
\begin{equation}
\kappa_T = - \frac{1}{V} \left( \frac{\partial P}{\partial V}
\right)_T = \frac{V}{N^2} \left( \frac{\partial N}{\partial \mu}
\right)_T.
\end{equation}

\begin{figure} [htb]
\centering
\begin{tabular}{cc}
\epsfig{file=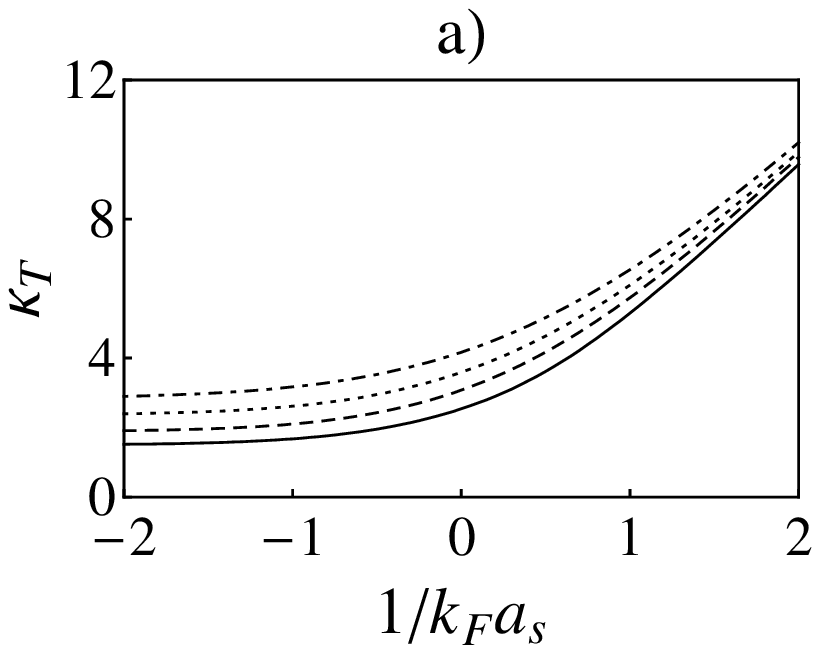,width=0.48 \linewidth} &
\epsfig{file=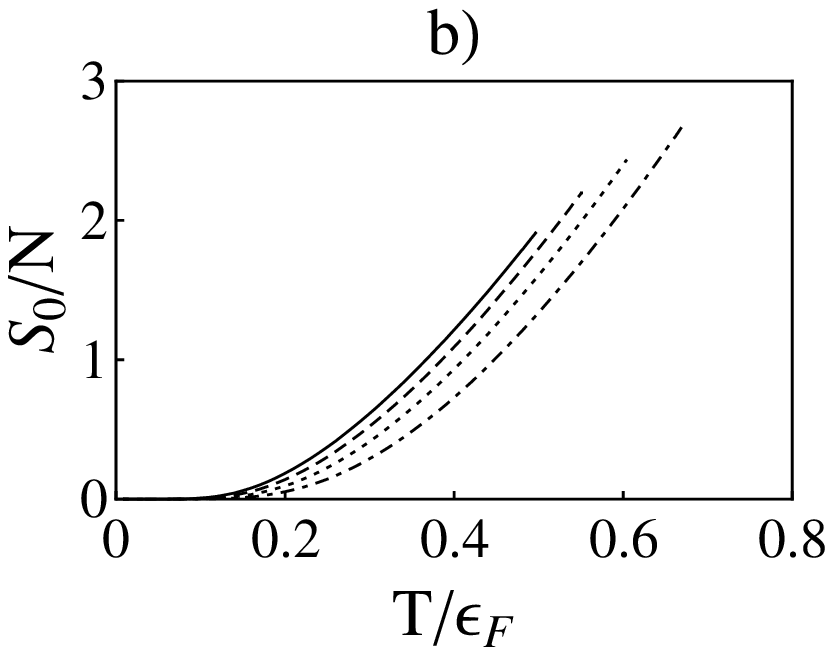,width=0.48 \linewidth}
\end{tabular}
\caption{ \label{fig:two} a) Compressibility $\kappa_T$ (in units of
$1/(n \epsilon_F)$) as a function of interaction parameter $1/(k_F
a_s)$ at $T = 0$ and b) entropy per particle $S_0/N$ as a function
of temperature $T$ (in units of $\epsilon_F$) at unitarity in the RO
case, for the same group of spin-orbit coupling values $v_R/v_F$ as
in Fig.~\ref{fig:one}. }
\end{figure}

As shown in Fig.~\ref{fig:two}a, for the RO case, the isothermal
compressibility $\kappa_T$ at fixed interaction parameter $1/( k_F
a_s )$ increases with increasing spin-orbit coupling $v_R$, as the
Fermi system becomes less degenerate reducing the Pauli pressure,
and thus more compressible. However, in the ERD case, the isothermal
compressibility for fixed interaction parameter does not change with
increasing spin-orbit coupling $v$. In this high symmetry situation
the momentum shift in the energy spectrum and the accompanied shift
in the chemical potential do not affect the degeneracy of the Fermi
system or the Pauli pressure, leading to an isothermal
compressibility which is independent of the spin-orbit coupling $v$.

%

%
%

{\it Equation of State and Entropy:}
Since the thermodynamic potential $\Omega = -PV$,
the saddle point pressure is
%
$
P_0 (T, \mu_{\alpha})
=
-
\Omega_0/V,
$
%
which can be shown to be always positive for arbitrary spin-orbit
coupling. The general trend of the pressure for fixed interaction
parameter (from the BCS to the unitarity regimes) is to decrease
with increasing spin-orbit coupling for both ERD and RO cases. The
situation in the BEC regime requires the inclusion of quantum
fluctuations to recover the corresponding Lee-Yang corrections in
the presence of spin-orbit effects. The entropy is then calculated
from $ S = -\left(
\partial \Omega /\partial T
\right)_{V, \mu_{\alpha}}.
$
In Fig.~\ref{fig:two}b, we show the saddle point entropy $S_0$
for the RO case at unitarity. For fixed $T$, $S_0$ decreases
with increasing spin-orbit coupling due to the stabilization
of superfluidity by the spin-orbit field.

%
%

{\it Spin Susceptibility Tensor:}
A rotation of the matrix ${\bf H}_0$
into the helicity
basis $\vert{\Uparrow\rangle}, \vert{\Downarrow\rangle}$
introduces order parameters $\Delta_{0, \Uparrow \Uparrow}$
and $\Delta_{0, \Downarrow \Downarrow}$,
which are controlled by the spin-orbit coupling.
The emergence of the triplet component affects
dramatically the spin susceptibility
of the system.
Using standard linear response theory~\cite{gorkov-2001},
the uniform spin susceptibility tensor per unit volume is
\begin{equation}
\chi_{ij}
=
-
\frac{\mu_B^2}{V}
\sum_{\bf k}
\left[
a_{ij} ({\bf k})
-
b_{ij} ({\bf k})
\right],
\end{equation}
where the spin-spin correlations in the single-particle channel are
$ a_{ij} ({\bf k}) = \sum_{i\omega} {\rm Tr} \left[ \sigma_i {\bf G}
({\bf k}, i\omega) \sigma_j {\bf G} ({\bf k}, i\omega) \right] $ and
in the pair (anomalous) channel are $ b_{ij} ({\bf k}) =
\sum_{i\omega} {\rm Tr} \left[ \sigma_i {\bf F} ({\bf k}, i\omega)
\sigma_j^T {\bf F}^\dagger ({\bf k}, i\omega) \right]. $ The
matrices $\mathbf{G}$ and $\mathbf{F}$ are the block matrices
appearing in the inverse of ${\bf M}$ defined in
Eq.~(\ref{eqn:matrix-m}),
$$
\widetilde{\bf M}^{-1} ({\bf k}, i\omega)
=
\left(
\begin{array}{cc}
{\bf G} & {\bf F} \\
{\bf F}^\dagger & {\overline {\bf G}} \\
\end{array}
\right).
$$

\begin{figure} [htb]
\centering
\begin{tabular}{cc}
\epsfig{file=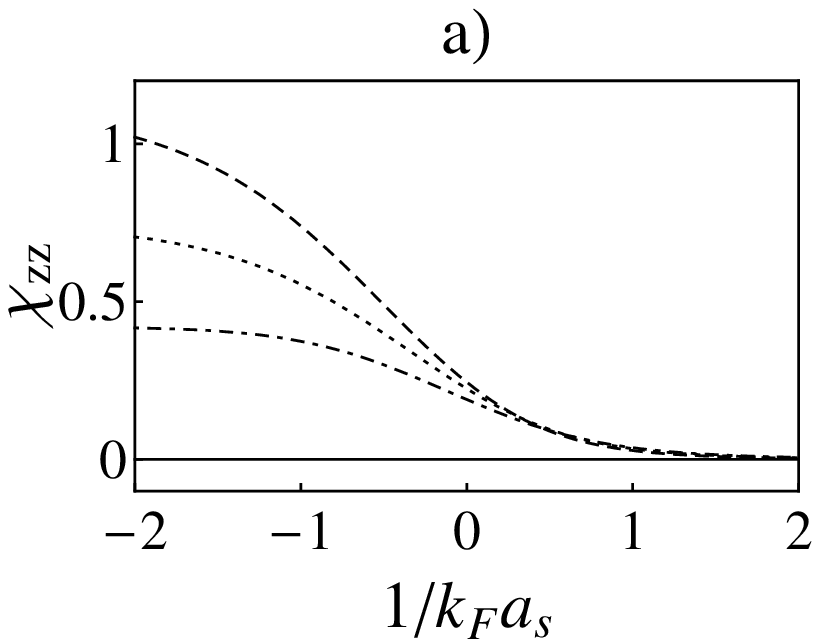,width=0.48 \linewidth} &
\epsfig{file=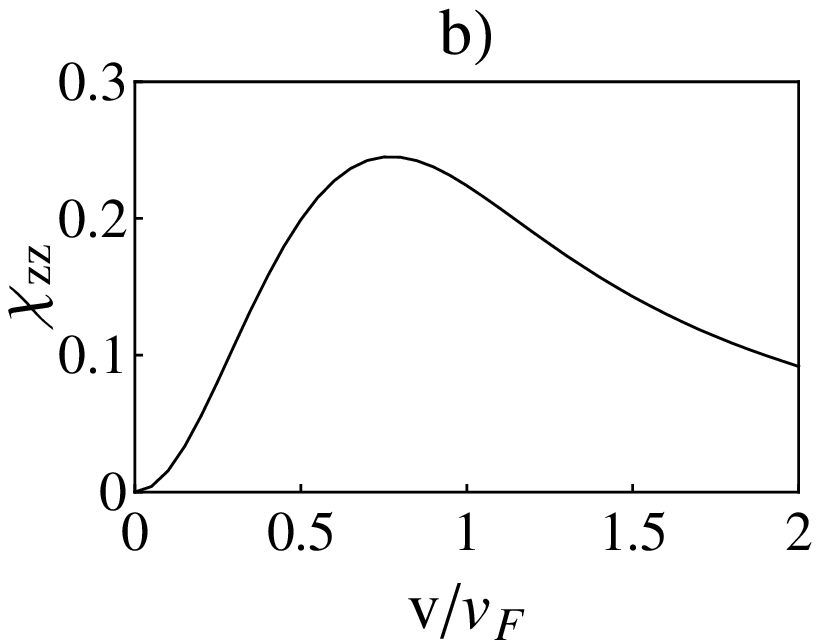,width=0.48 \linewidth}
\end{tabular}
\caption{ \label{fig:three} a) Spin susceptibility $\chi_{zz}$ (in
units of $\mu_B^2 n/\epsilon_F$) as a function of $1/(k_F a_s)$ at
$T = 0$ in the ERD case for $v/v_F = 0$ (solid), $v/v_F = 0.8$
(dashed), $v/v_F = 1$ (dotted), and $v/v_F = 1.2$ (dot-dashed). b)
Spin susceptibility $\chi_{zz}$ as a function of $v/v_F$ at $T = 0$
at unitarity in the ERD case. }

\end{figure}

In Fig.~\ref{fig:three}a, we show plots of $\chi_{zz}$ for the ERD
case at $T = 0$ as a function of $1/(k_F a_s)$ for various values of
spin-orbit coupling, and the behavior of $\chi_{zz}$ for the RO case
is qualitatively similar. In Fig.~\ref{fig:three}b, we show
$\chi_{zz}$ versus $v$ in the unitary limit $1/(k_F a_s) = 0$. The
maximum in $\chi_{zz}$ corresponds to the maximum in the triplet
component of $\Delta_0$. For small and large $v$ the triplet
component is small.

In the ERD case $\chi_{zz} = \chi_{xx} \ne \chi_{yy}$, and in the
zero temperature limit $\chi_{yy} (T \to 0) = 0$, while $\chi_{zz} =
\chi_{xx}$ remains finite for non-zero spin-orbit coupling. In the
RO case $\chi_{zz} \ne \chi_{xx} = \chi_{yy}$, and in the $T \to 0$
limit $\chi_{xx} (T \to 0) = \chi_{yy} (T \to 0) = \chi_{zz} (T \to
0)/2$. Lastly, for $h_\perp(\mathbf{k}) = 0$ (no spin-orbit
coupling) the spin susceptibilty tensor becomes $\chi_{ij} = \chi
\delta_{ij}$, where the scalar $ \chi = \left[ \mu_B^2/(2VT) \right]
\sum_{\bf k} {\rm sech}^2 \left[ \sqrt{\xi_{\bf k}^2 + \vert
\Delta_0 \vert^2} / (2T) \right] $ is the Yoshida function, which
vanishes at zero temperature, i.e., $\chi (T \to 0) = 0$. The
existence of non-zero spin response even at $T = 0 $ is a direct
measure of the induced triplet component of the order parameter due
to the presence of spin-orbit coupling, since that a pure singlet
superfluid at $T = 0$ must have zero spin susceptibility since all
fermions are paired into a zero-spin state.

{\it Conclusions:} We have studied the effects of spin-orbit
coupling in the evolution from BCS to BEC superfluidity at low
temperatures, and concluded that this evolution is just a crossover.
We discussed effects of spin-orbit coupling on thermodynamic
properties including the order parameter, chemical potential,
pressure, entropy, isothermal compressibility and spin
susceptibility tensor to support the crossover picture. We also
proposed way to experimentally detect the spin-orbit coupling effect
by measuring the spin susceptibility.

\acknowledgements{We would like to thank NSF (Grant No. DMR-0709584)
and ARO (Contract No. W911NF-09-1-0220).}

\end{document}